\documentclass[a4paper]{panle}
\usepackage{url}
\usepackage{cite}
\usepackage{wrapfig}
\usepackage{graphicx}
\usepackage{amssymb}
\usepackage{amsfonts}
\usepackage{amsmath}
\usepackage{longtable}
\usepackage{rotating}
\usepackage{lscape}
\usepackage{epsfig}
\usepackage{multirow}
\usepackage{bm}

\originalTeX

\begin{document}

\title{A method for multimodal analysis of TAIGA experiment data using essential features}
\maketitle
\authors{A.\,P.~Kryukov\,$^{a,b,}$\footnote{E-mail: kryukov@theory.sinp.msu.ru}, 
J.\,J.~Dubenskaya\,$^{a}$, 
E.\,Yu.~Fedotova\,$^{a}$, 
E.\,O.~Gres\,$^{a,b}$, 
S.\,P.~Polyakov\,$^{a}$ 
E.\,B.~Postnikov\,$^{a}$ 
A.\,Y.~Razumov\,$^{a}$ 
P.\,A.~Volchugov\,$^{a}$, 
D.\,P.~Zhurov\,$^{a,b}$
}
\setcounter{footnote}{0}
\from{$^{a}$\,Lomonosov Moscow State University, Skobeltsyn Institute of Nuclear Physics, Moscow, 119991 Russia}
\from{$^{b}$\,Research Institute of Applied Physics, Irkutsk State University, Irkutsk, 664003 Russia}


\begin{abstract}

The aim of processing and analyzing experimental data from physical experiments is to obtain physically significant information about the phenomenon under study. This goal is achieved by multi-stage processing of experimental data, during which noise associated with measurements is suppressed and the dimensionality of the input data is reduced. 

In this paper, we propose a new method based on the use of neural networks such as autoencoders to extract essential features. The special value of the proposed approach lies in the possibility of its application to the analysis of multimodal data received simultaneously from several installations.

We will apply this approach to a multimodal data (MMD) of the experiment TAIGA. 
Currently, the analysis of the MMD is carried out independently for each installation separately. Therefore, the development of methods for the joint analysis of MMD from TAIGA-type installations is an urgent task in cosmic ray physics and gamma-ray astronomy. 

Based on Monte Carlo simulation, it is shown that the proposed method allows for effective MMD analysis. It can also be used for MMD analysis at other experimental complexes.
\end{abstract}
\vspace*{6pt}


\label{sec:intro}
\section*{INTRODUCTION}

The objective of any physical experiment is to obtain physically significant information about the phenomenon under study. However, the complexity of both the physical phenomenon itself and the process of its registration requires multi-stage processing of experimental data, during which noise associated with measurements is suppressed and the dimensionality of the input data is reduced. In general, this process can be called the process of extracting essential features (EF). The final set of these essential features is used to obtain parameters that can be interpreted within the framework of physical models of the phenomenon under study. 

Traditionally, a set of essential features extracted from experimental data are physically motivated parameters, since this corresponds to the logic of scientific investigation. For machine learning methods~\cite{Demichev:Astro2024} this requirement is optional. Moreover, neural networks are able to work with data representations that either do not have a direct interpretation, or their interpretation is difficult.

Convolutional autoencoders can be used to extract essential features from image data~\cite{Masci:ANN2011,Li:ASC2023,Yao:ICPHM2019,Meng:IJCNN2017}. 
In this case, a latent space of the AE bottle neck can be considered as a space of essential features. 
Since the dimension of the AE lattent space is much smaller than the dimension of the input data, and the image reconstructed from it is close to the original one, it is suggest that the representation of the data in the latent space still contains all  necessary physical information. 
Moreover, the property of AE to suppress noise also helps in solving our task. 
The main advantage of using AE in this way is the ability to train a neural network without prior marking up the training sample. 

The proposed method was tested on simulated data for TAIGA installations~\cite{Budnev:PPN2025}. 
The TAIGA complex includes several atmospheric Cherenkov telescopes (IACT) and an array of wide-angle Cherenkov HiSCORE detectors. 
Currently, the analysis of this multimodal data (MMD) is carried out independently for each installation. 
Therefore, the development of methods for the joint analysis of MMD from TAIGA-type installations is an urgent task in cosmic ray physics and gamma-ray astronomy. 

This paper demonstrates the use of AE and the essential features extracted with its help to restore the parameters of EAS. 
The results obtained are compared with traditional methods of analysis. It has been shown that in some cases, this method provides a significant increase in the accuracy of determining EAS parameters.
The special value of the proposed approach lies in the possibility of its application to the analysis of multimodal data received simultaneously from several installations.

This work is a continuation of the work of \cite{Gres:PAN2025}.

\label{sec:taiga}
\section{Unique experimental complex TAIGA}


The Unique experimental complex TAIGA~\cite{Budnev:PPN2025} consist of a some experimental facilities. The main of them are four Imaging atmospheric Cherenkov telescopes, the array of Cherenkov detector station HiSCORE.

IACTs detect Cherenkov light emitted by secondary particles of EAS.
The each IACT contain a camera with about 600 PMTs assembled to hexagonal grid. Each PMT record the amplitude of input signal. 

HiSCORE
 is an array of wide-angle Cherenkov detector stations with a wide Field of View (FoV ~0.6 sr). 
The detector stations are placed at about 150m from each other. 
The more then 120 stations of the array cover an area of ~1 km². 
Each station records amplitude of input signal, duration of them and time of arrival.

Thus, the IACTs recording a distribution of Cherenkov light from EAS.
And HiSCORE records a spatio-temporal distribution of the Cherenkov light from EAS at the Earth level. 

The traditional EAS data processing method includes two stages. First, using HiSCORE data, the arrival point of the EAS and the direction of its axis are determined. 
In the second step, the IACT image data convert to the Hillas parameters. 
Obtained Hillas parameters use to determinate the energy of the EAS and the type of primary particle together with parameters from HiSCORE.

Thus, the traditional method uses a two-stage analysis, which does not use combining data from both experimental sets for joint analysis.

\label{sec:mmd}
\section{Multimodal data analysis of TAIGA facilities}

The joint analysis of data from several types of installations, for example, as in our case, IACT and HiSCORE, involves compressing the input data to a quantity of parameters that ensure acceptable accuracy of restoring the input data using autoencoders. In our case, we have determined that the optimal dimension of the hidden space of autoencoders is 12 (see Fig.~\ref{fig01}).

The realistic neural network that was used is shown in Fig.~\ref{fig02}. Here, data in the hidden space of autoencoders for IACT and geometric parameters extracted from HiSCORE data using 1D convolution networks were used as essential features.
All these essential features were concatenated into a single vector, which was transfer to a fully connected (multitask) neural network to extract the physical parameters of the EAS.


\begin{figure}[t]
\begin{center}
\includegraphics[width=120mm]{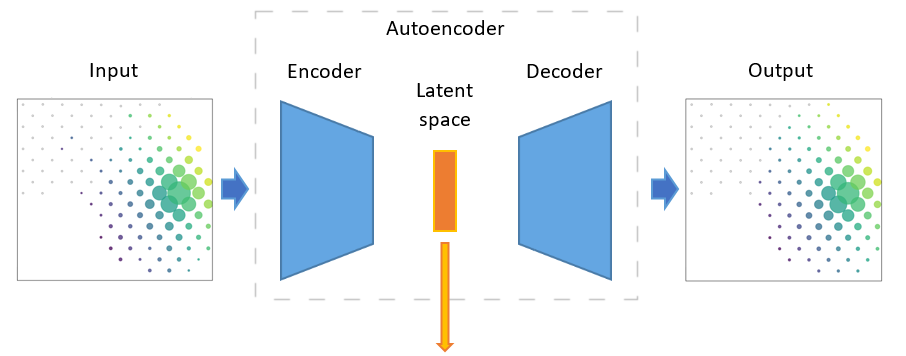}
\vspace{-3mm}
\caption{Feature extraction using autoencoder. Latent space as a space of essensial features.}
\end{center}
\labelf{fig01}
\vspace{-5mm}
\end{figure}

\begin{figure}[t]
	\centerline{\includegraphics[width=0.5\textwidth]{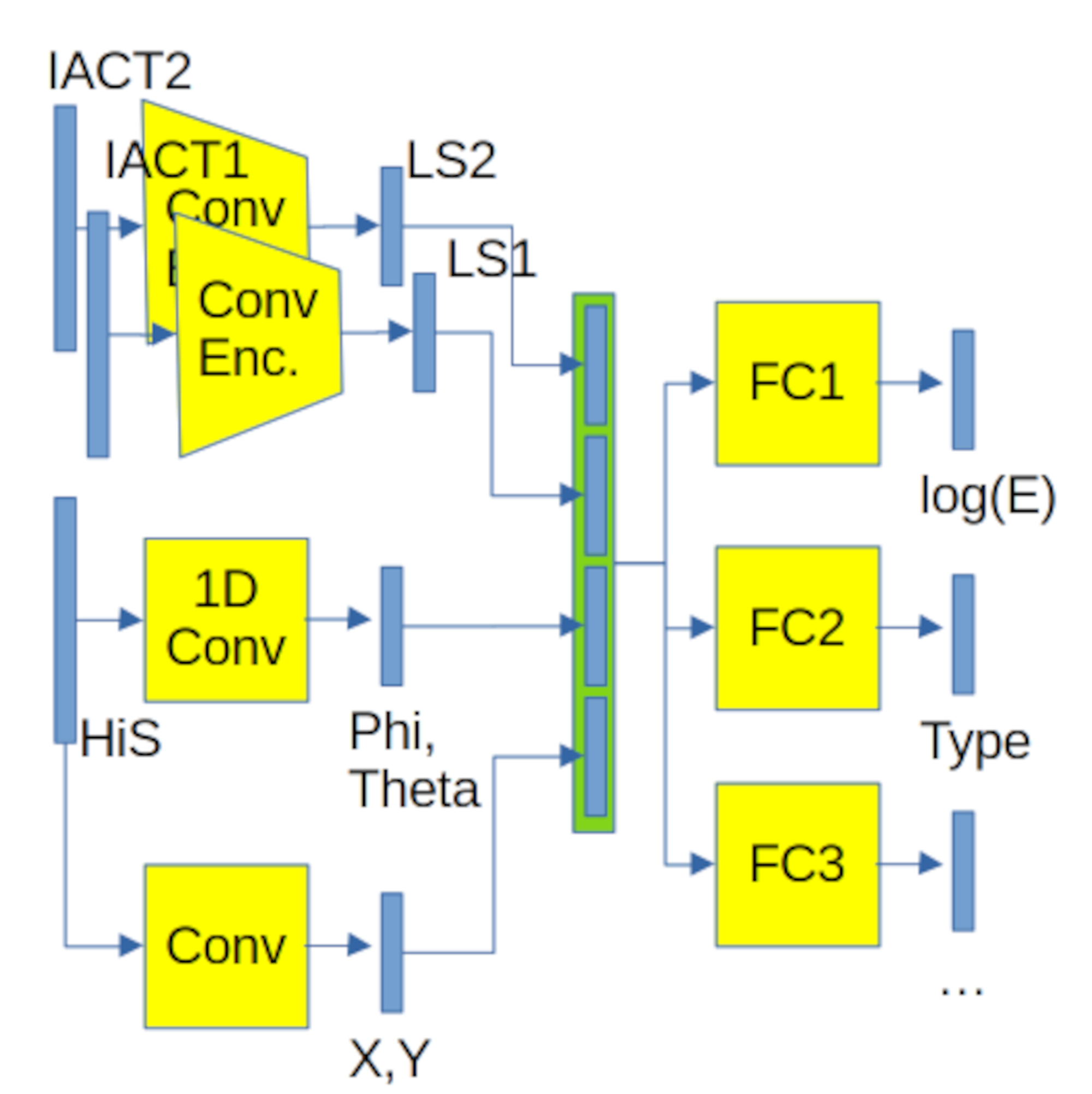}}
		\vspace{-3mm}
		\caption{Realistic scheme for multimode data analysis of IAVT and HiSCORE data.}
	\labelf{fig02}
\end{figure}

\section{Discussion and results}


The proposed method has been tested on a fairly wide class of tasks. 
Here we will focus on the problem of determining the point of arrival of SHAL and the energy of SHAL. The obtained results were compared with the data obtained by traditional methods on the same test sample.

Neural network training was performed on Monte Carlo events that were combined for IACT1 and HiSCORE. The approximate number of such events was about 30,000. About 10,000 more joint events were used as a validation set. The final comparison of the results obtained using neural networks and the traditional method was carried out on a test set of about 2000 events that were not present in the training and validation sets. All events were generated in the energy range from 40 to 400 TeV.

Figure~\ref{fig04} shows the distribution of the absolute error of the reconstructed EAS energy by two different neural networks (blue and red histogram) and the traditional method (green histogram). Note that the traditional method underestimates the energy of EAS, which results in a systematic error in determining the energy of 56 TeV. The relative error of the reconstruction of the traditional method decreases from 0.8 at 40 TeV to 0.3 at 400 TeV. The relative error of machine learning reconstruction decreases from 0.3 at 40 TeV to 0.1 at 400 TeV.

\begin{figure}[t]
\centerline{\includegraphics[width=120mm]{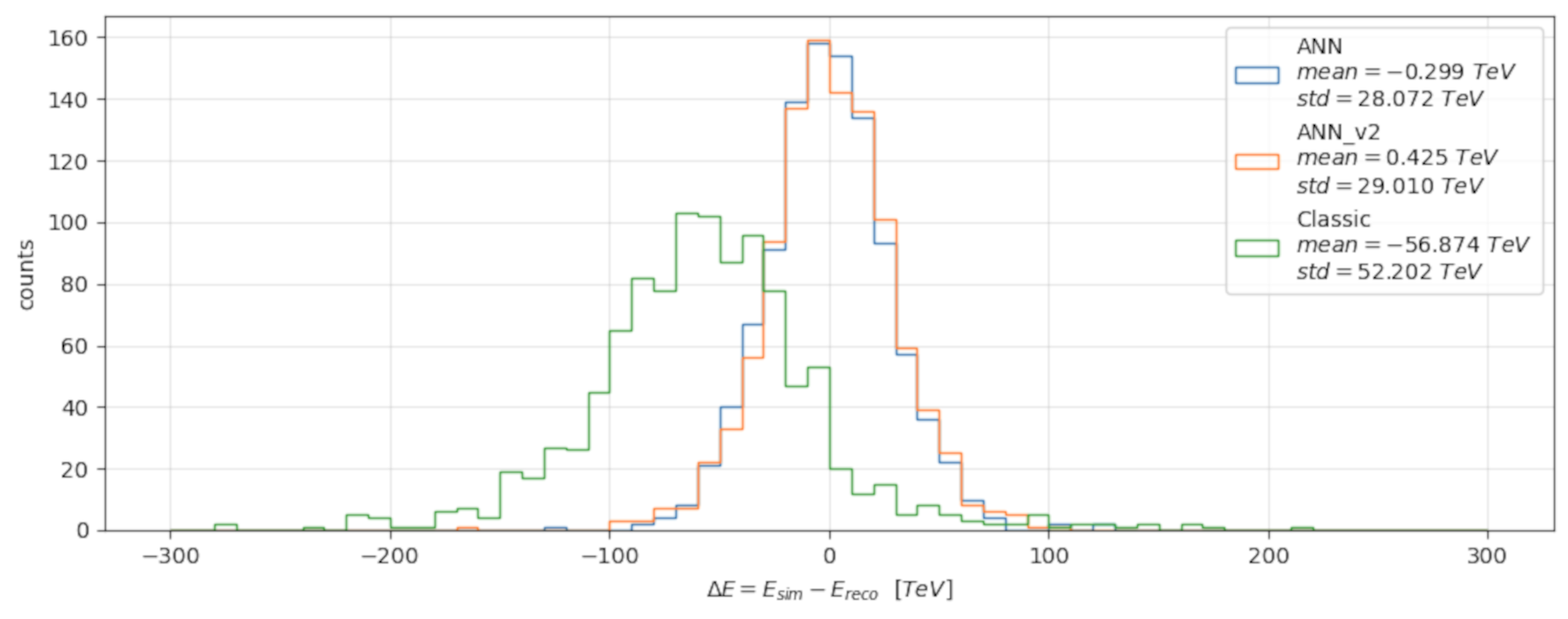}}
		\vspace{-3mm}
		\caption{Absolute energy error distribution. Comparison of traditional method and proposed method.}
	\labelf{fig04}
	\vspace{-5mm}
\end{figure}

A similar comparison was carried out for the task of determining the point of arrival of the EAS axis. For the EAS core, both approaches do not provide significant shifts of core position. The shift was less than a meter for the neural networks and a little bit more then 1 meter for the traditional methods. The error in determining the position of the EAS core for the machine learning method was about 6 m, which is more than 2 times less than the analog result obtained by the traditional method. Figure~\ref{fig03} shows the distribution of the deviation of the reconstructed EAS arrival points from the true position according to Monte Carlo data. The figure clearly shows the advantage of the proposed method over the traditional approach.

\begin{figure}[t]
\centerline{\includegraphics[width=1\textwidth]{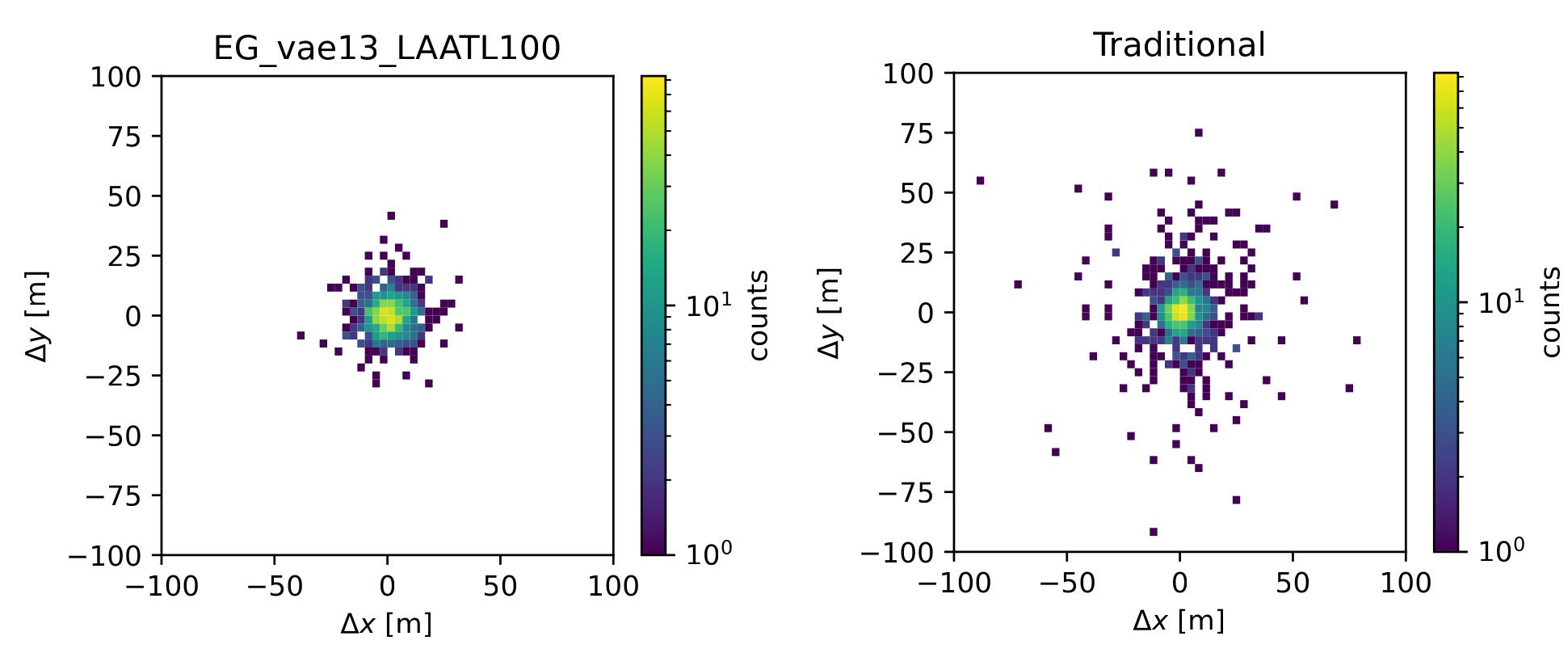}}
		\vspace{-3mm}
		\caption{Distribution of errors of EAS core. Comparison of traditional method and proposed method.}
	\labelf{fig03}
	\vspace{-5mm}
\end{figure}

\section*{CONCLUSIONS}

Machine learning is a powerful method for analyzing data from ground-based cosmic-ray and gamma-ray astronomy facilities. Using autoencoder-type neural networks for feature extraction makes it possible to naturally combine data from multiple facilities of different types at the essential feature level and to make a joint (multimodal) analysis.

The advantage of this approach is that the autoencoder's latent space is treated as an abstract representation of the input data, eliminating the need to interpret these parameters in physically motivated terms.

Using a multitask, fully connected neural network for the final analysis and extraction of extensive air shower (EAS) physical parameters allows for the training of specific network "heads" adapted to the characteristics of the physical parameters being reconstructed.

Computational experiments conducted on simulated Monte Carlo data demonstrated a significant advantage of the proposed method over the traditional data analysis approach. The accuracy of determining a number of extensive air shower (EAS) parameters is two to three times higher than that of traditional methods.

\section*{FUNDING}

The study was conducted as part of a state assignment to 
Lomonosov Moscow State University using equipment provided by 
the MSU Development Program and data obtained from 
the unique TAIGA research facility.

\section*{ACKNOWLEDGEMENTS}

The author expresses his gratitude to the TAIGA collaboration for the useful discussions and the model data provided. The author is also grateful to E.~Osipova for her help in comparing the proposed method with the traditional one.

\section*{CONFLICT OF INTEREST}

The authors declare that they have no conflicts of interest.

\bibliographystyle{pepan}
\bibliography{K12_biblio}

\end{document}